\documentclass[preprint,aps,prb]{revtex4-1}

\usepackage{amssymb}
\usepackage{amsmath}
\usepackage{graphicx}
\usepackage{epstopdf}
\usepackage{color}
\usepackage{subfigure}
\usepackage{hyperref}
\usepackage{xcolor}
\usepackage [english]{babel}
\usepackage [autostyle, english = american]{csquotes}
\MakeOuterQuote{"}

\begin{document}

\preprint{}

\title{Structural and Magnetic Phase Transitions in Chromium Nitride Thin Films Grown by RF Nitrogen Plasma Molecular Beam Epitaxy}

\author{{Khan Alam$^{1}$, Steven M. Disseler$^{2}$, William D. Ratcliff$^{2}$, Julie A. Borchers$^{2}$, Rodrigo Ponce-P\'{e}rez$^{1,3}$, Gregorio H. Cocoletzi$^{3}$, Noboru Takeuchi$^{1,4}$, Andrew Foley$^{1}$, Andrea Richard$^{1}$, David C. Ingram$^{1}$, and Arthur R. Smith$^{1,*}$}\footnote[0]{*E-mail: smitha2@ohio.edu, Fax: 740-593-0433}}

\affiliation{$^{1}$Nanoscale and Quantum Phenomena Institute - Department of Physics and Astronomy, Ohio University, Athens, OH 45701, USA}

\affiliation{$^{2}$NIST Center for Neutron Research - National Institute of Standards and Technology, Gaithersburg, MD 20899, USA}

\affiliation{$^{3}$Instituto de F\'{i}sica, Benem\'{e}rita Universidad Aut\'{o}noma de Puebla, Apartado Postal J-48, Puebla, M\'{e}xico}

\affiliation{$^{4}$Centro de Nanociencias y Nanotecnolog\'{i}a, Universidad Nacional Aut\'{o}noma de M\'{e}xico,  Ensenada Baja California, Código Postal 22800, M\'{e}xico}

%%%%%%%%%%%%%%%%%%%%%%%%%%%%%%%%%%%%%%%%%%%%%%%%%%%%

\vskip 1cm
\begin{abstract}

A magneto-structural phase transition is investigated in single crystal CrN thin films grown by rf plasma molecular beam epitaxy on MgO(001) substrates. While still within the vacuum environment following MBE growth, {\it in-situ} low-temperature scanning tunneling microscopy, and {\it in-situ} variable low-temperature reflection high energy electron diffraction are applied, revealing an atomically smooth and metallic CrN(001) surface, and an {\it in-plane} structural transition from 1$\times$1 (primitive CrN unit cell) to $\mathrm{\sqrt{2}\times\sqrt{2}-R45^\circ}$ with a transition temperature of $\sim$ 278 K, respectively. {\it Ex-situ} temperature dependent measurements are also performed, including x-ray diffraction and neutron diffraction, looking at the structural peaks and likewise revealing a first-order structural transition along both [001] and [111] {\it out-of-plane} directions, with transition temperatures of 256 K and 268 K, respectively. Turning to the magnetic peaks, neutron diffraction confirms a clear magnetic transition from paramagnetic at room temperature to antiferromagnetic at low temperatures with a sharp, first-order phase transition and a N\'{e}el temperature of 270 K or 280 K for two different films. In addition to the experimental measurements of structural and magnetic ordering, we also discuss results from first-principles theoretical calculations which explore various possible magneto-structural models.


\end{abstract}

\maketitle
\newpage
\normalsize

%%%%%%%%%%%%%%%%%%%%%%%%%%%%%%%%%%%%%%%%%%%%%%%%
\newpage
\section{Introduction}

Although originally known for its impressive physical properties including high hardness and corrosion resistance \cite{wiklund1997cracking, nouveau2001stress}, chromium nitride (CrN) has attracted considerable attention in recent years due to its potential use as an electronic or spintronic material resulting from observed semiconducting-like behavior over a variety of temperatures\cite{gall2002band}, and its antiferromagnetic ordering at low temperatures\cite{anderson2005magnetic}. Besides potential electronic applications, CrN could be a model system for studying first-order phase transitions in which structural, electronic, and magnetic properties are intertwined.

The structural, electronic, and magnetic properties of CrN were originally understood based only on results from bulk powder samples \cite{corliss1960antiferromagnetic, bhobe2010evidence}. For example, it was known since 1960 that bulk CrN is paramagnetic (PM) with a rock-salt crystal structure at room temperature (RT), but becomes antiferromagnetic (AFM) at low temperature (LT), with a N\'{e}el temperature T$_{N}$ of 273-283 K and with an orthorhombic crystal structure \cite{corliss1960antiferromagnetic, mrozinska1979first}. Then Filippetti {\it et al.} proposed in 2000 that magnetic stress is the driving force for the structural transition,\cite{filippetti2000magnetic} thus linking the magnetic with the structural.

While the properties of CrN seemed quite well understood for bulk material by 2000, since then various groups have reported highly discrepant properties for CrN in thin film form. For example, Gall \textit{et al.} reported that CrN thin films grown on MgO(001) by reactive magnetron sputtering, due to epitaxial constraints, do not exhibit a structural transition and that the resistivity varies by orders of magnitude, increasing with decreasing temperature\cite{gall2002band}. Whereas in 2004, Constantin {\it et al.} found instead, semiconducting behavior at room temperature but a clear transition to metallic behavior at low temperature for CrN thin films grown by radio-frequency nitrogen plasma molecular beam epitaxy (rf N-plasma MBE); however, Constantin {\it et al.} did not address the presence or absence of either structural or magnetic transitions in their films \cite{constantin2004metal}.

In 2011, Zhang \textit{et al.} reported the absence of a structural transition for single crystal CrN thin films grown on MgO(001) and MgO(111) substrates by sputtering, but they did observe a transition for a polycrystalline CrN film grown on quartz\cite{zhang2011epitaxial}. Yet again, Inumaru \textit{et al.} studied CrN/MgO(001) and CrN/sapphire(0001) thin films grown by pulsed laser deposition, and while they did observe a structural transition for CrN/MgO(001), they did not for the case of CrN/sapphire(0001) \cite{inumaru2007controlling}.

Ney \textit{et al.} investigated the magnetic properties of CrN thin films grown by molecular beam epitaxy (MBE) on two different substrates, MgO(001) and sapphire(0001). They reported that CrN/MgO(001) showed no magnetic transition from PM to AFM upon cooling, although CrN/sapphire(0001) showed ferromagnetic-like behavior at low temperatures\cite{ney2006magnetic}.

Herwadkar and Lambrecht's 2009 paper examined and attempted to address some of the discrepant reports by means of electronic structure calculations using the local spin density approximation including Hubbard correction (LSDA+U) applied to the Corliss AFM model (referred to as AFM-[110]$_2$) as well as some competing models. They proposed possible reasons for the widely differing transport properties reported, and the observation or lack thereof, of structural/magnetic phase transitions suggesting, for example, that the various properties could be strongly affected by the presence/abscence of N vacancies and possible localization effects \cite{herwadkar2009electronic}.

In this paper, we apply variable temperature reflection high energy electron diffraction (VT-RHEED), variable temperature x-ray diffraction (VT-XRD), and variable temperature neutron diffraction (VT-ND) to investigate a possible structural phase transition, and variable temperature neutron diffraction to investigate a possible magnetic phase transition in CrN thin films grown using rf N-plasma MBE. The experimental results may be compared to, and are consistent with, the temperature-dependent resistivity results reported by Constantin \textit{et al.} in 2004 \cite{constantin2004metal}. We also carry out first-principles theoretical calculations for CrN using several different computational methods in order to investigate structural and magnetic models which are consistent with our experimental results. Although we will show that the measurements do not uniquely support one particular structural model, the structural phase transition is very clear, both for the {\it in-plane} as well as {\it out-of-plane} measurements, although there are some variations in the measured structural transition temperatures. On the other hand, we will also show that the magnetic measurements uniquely support the magnetic model (AFM-[110]$_2$) put forth by Corliss {\it et al.} in 1960.\cite{corliss1960antiferromagnetic} We also find that sample stoichiometry (nitrogen deficiency) appears to affect the transition temperature slightly.

%%%%%%%%%%%%%%%%%%%%%%%%%%%%%%%%%%%%%%%%%%%%%%%%%%%%%
\section{Methods}

Growth and VT-RHEED experiments are performed in a custom designed ultra-high vacuum (UHV) system combining MBE with low temperature spin-polarized scanning tunneling microscopy (LT-SP-STM), which makes excellent conditions for \textit{in situ} characterization of as-grown thin films \cite{lin2014facility}. High quality CrN thin films are grown on MgO(001) substrates initially cleaned \textit{ex situ} with acetone followed by isopropyl-alcohol and further prepared in vacuum by heating up to 1000$\pm$30 $^{^\circ}$C while being exposed to nitrogen plasma flux until a streaky RHEED pattern is obtained. All four samples (film thicknesses are given in the parentheses) S45 (670 nm), S61 (980 nm), S73 (150 nm), and S75 (37 nm) investigated in this experiment are grown under nitrogen rich conditions at 650$\pm$30 $^{^\circ}$C without any buffer layer except for S45, where a 21 nm CrN buffer layer was grown at 400$\pm$20 $^{^\circ}$C. The Cr flux is obtained by heating a carefully degassed effusion cell, and nitrogen flux (1.38$\times$10$^{14}$ atoms/cm$^{2}$s) is obtained from a radio frequency N-plasma generator, which uses 99.999\% ultra high pure nitrogen gas source and operates at 450 W forward power. The growths are controlled by opening and closing the Cr effusion cell shutter, and the entire growth process is monitored with RHEED. The Cr flux is measured using quartz crystal thickness monitor, and N flux is calibrated by growing a 1$\times$1 GaN film \cite{alam2015native}. During S45 growth, Cr flux is varied in a range of 2-5$\times$10$^{13}$  atoms/cm$^{2}$s whereas the flux is kept constant at 2 $\pm$ 0.2$\times$10$^{13}$ atoms/cm$^{2}$s during S61, S73, and S75.

Chromium nitride samples are further studied to optimize growth conditions for high quality films using a wide variety of surface and bulk sensitive techniques including LT-STM, Rutherford back-scattering spectrometery (RBS), and x-ray diffraction (XRD).

A possible \textit{in-plane} structural transition is studied \textit{in situ} in CrN samples S73 and S75 using VT-RHEED. The possible \textit{out-of-plane} structural transition is studied \textit{ex situ} in S45 and S61 using VT-XRD, and VT-ND, respectively. The VT-RHEED setup consists of a custom designed VT sample stage, which can be heated up to 1273 K and cooled down to 193 K and a RHEED system for continuously monitoring surfaces. We used liquid nitrogen to cool down the samples while observing changes in the surfaces with a 20 keV e-beam. The experiments are performed one at a time for four crystallographic directions: [100], [110], [130], and [$\underline{1}$20], and the patterns are recorded above and below the transition temperature. In the VT-XRD experiment nitrogen vapor is used to cool down the sample. X-ray spectra of the 002 peak of MgO and CrN are continuously recorded as the sample is either cooling down or warming up. The CrN temperature is determined by using lattice constant and thermal expansion coefficient of MgO in the linear thermal expansion equation.

We performed variable temperature neutron diffraction (VT-ND) experiments at the NIST Center for Neutron Research (NCNR). Neutron diffraction measurements were performed to study the magnetic and structural transitions in S45 and S61 using the BT-4 triple axis spectrometer with neutrons of incident energy 14.7 meV (corresponding to wavelength 2.359 \AA) for the magnetic reflections and 30.5 meV (corresponding to wavelength 1.638 \AA) for the structural reflections. Energies were selected with a pyrolitic graphite monochromator and analyzer and also pyrolitic graphite filters were placed in the beam to remove contamination from higher order wavelengths. To maximize signal intensity, 40$\prime$ collimations were used before and after the sample with open collimation before the monochromator and after the analyzer. Collimation settings for both neutron diffraction experiments were the same. Samples were mounted on single-crystal silicon wafers with fluorinated grease and sealed with He atmosphere inside an aluminum can for temperature control via a closed-cycle refrigerator. In addition to the BT-4 triple axis spectrometer data presented in this paper, polarized neutron beam experiments were also performed using the BT-7 triple-axis spectrometer with 14.7 meV neutrons and $^3$He neutron spin filters.\cite{lynn2012double,chen20073}

Calculations have been done within the periodic density functional theory (DFT) as implemented in the PWscf code of the Quantum ESPRESSO package \cite{giannozzi2009quantum}. Exchange-correlation energies are modeled using three different approximations: the local density approximation with the Perdew-Zunger (PZ) parametrization \cite{perdew1981self} plus Hubbard correction (LDA+U) \cite{anisimov1997first} with 3 eV $\leq$ U $\leq$ 5 eV employing the simplified version of Cococcioni \cite{cococcioni2005linear}; the generalized gradient approximation (GGA) as stated by the Perdew-Burke-Ernzerhof (PBE) gradient corrected functional \cite{perdew1996generalized} and the GGA-PBE functional plus Hubbard correction (GGA+U) with 1 eV $\leq$ U $\leq$ 5 eV using the Cococcioni simplified version \cite{cococcioni2005linear}. In all cases, Vanderbilt ultra-soft pseudopotentials \cite{vanderbilt1990soft} have been employed to replace the core electrons. The cut-off energy to truncate the electronic states expansion in plane waves has been optimized, finding a cutoff of 30 Ry to be appropriate. For the charge density, we have used a density cutoff of 240 Ry. Convergence was achieved when the forces acting on each ion were smaller than 0.002 eV/\AA, and the energy difference between two consecutive steps was less than 0.01 eV.  Moreover, Brillouin zone integration has been done using a Methfessel-Paxton \cite{methfessel1989high} smearing of 0.01 Ry and an optimized and equally spaced k-points mesh of 5x5x5 \cite{monkhorst1976special} centered at Gamma.

\section{Results and Discussion}

\subsection{Growth and {\it In-situ} Sample Characterization}

\begin{figure}[htbp]
\centering
\includegraphics[angle=0,width= 5 in]{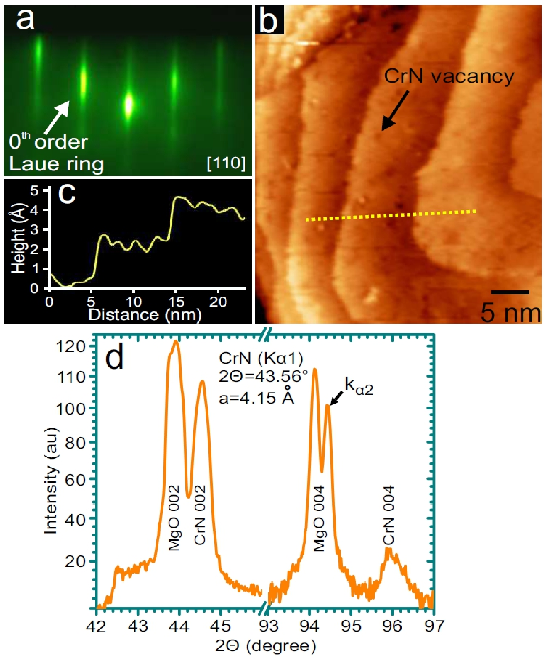}

\subfigure{\label{fig:fig1-a}}
\subfigure{\label{fig:fig1-b}}
\subfigure{\label{fig:fig1-c}}
\subfigure{\label{fig:fig1-d}}

\caption{ \subref{fig:fig1-a} Streaky RHEED pattern (for S45) recorded along [110] showing zeroth order Laue zone ring, corresponding to cubic symmetry of CrN at room temperature. \subref{fig:fig1-b} LT-STM image taken at LHe temperature (for S45) showing atomically smooth terraces with uniform contrast, except for some darker spots related to surface Cr (and/or N) vacancies. \subref{fig:fig1-c} Line profile taken from the dotted (yellow) line in the STM image showing a step height of 2.07 \AA. \subref{fig:fig1-d} XRD spectrum (for S45) showing 002 and 004 peaks of MgO and CrN; although hardly distinguished in the MgO and CrN 002 peaks, the MgO 004 shows both K$_{\alpha1}$ and K$_{\alpha2}$ peaks, but these are not distinguished for CrN 004.}
\label{fig:fig1}
\end{figure}

Shown in Fig.~\ref{fig:fig1-a} is a characteristic RHEED pattern of CrN, which is taken along [110] at the end of the growth of S45 at room temperature. Five sharp and continuous streaks in the zeroth-order Laue ring (Z$_{[110]}$) corresponding to the cubic symmetry of CrN are visible in this pattern. The streak spacing corresponds to an \textit{in-plane} lattice constant (a$_{\Vert}$) of 4.14 \AA, which shows that this sample is more relaxed than the Constantin \textit{et al.} samples (4.04-4.13 \AA) and matches the Ney \textit{et al.} sample (4.14 \AA) and can be taken as the bulk value\cite{constantin2004metal,ney2006magnetic}. The bulk lattice parameter of MgO is 4.213 \AA.\cite{yang2002crystalline}

A 38 nm $\times$ 42 nm LT-STM image of S45 taken at 4.2 K is presented in Fig.~\ref{fig:fig1-b}. Five atomically smooth terraces with consistent contrast are visible in the image. As well, some visible darker spots in all terraces corresponding to CrN surface vacancies can be seen. All terraces do not show any long-range-topographic distortions (LTDs) associated with semiconducting behavior, which were observed using STM by Constantin \textit{et al.} in CrN thin films at room temperature \cite{constantin2004metal}. The absence of LTDs indicates a metallic nature of our sample at low temperatures.

In order to find the \textit{out-of-plane} lattice constant (a$_{\bot}$), a line profile is taken across three terraces in the STM image as shown in Fig.~\ref{fig:fig1-c}. The step height of each terrace is 2.07 \AA, which is half of the a$_{\bot}$ of CrN.

\subsection{{\it Ex-situ} Sample Characterization}

Shown in Fig.~\ref{fig:fig1-d} is an XRD pattern of S45, where 002 and 004 peaks of CrN and the same peaks of MgO can be seen. Scattering of K$_{\alpha1}$ x-rays coming from the Cu target produce the main peaks and K$_{\alpha2}$ appears as a bulge on the side of the MgO 002 peak and as a second peak on the right side of the MgO 004 reflection. The XRD pattern is first calibrated with respect to the MgO substrate, and then the lattice parameter of CrN is measured. The 002 peak of CrN occurs at 2$\theta$ = 43.56$^{^\circ}$ corresponding to a$_{\bot}$ = 4.15 \AA, and it is in the range of the previously reported values of 4.13-4.17 \AA \cite{ney2006magnetic, constantin2004metal, corliss1960antiferromagnetic, gall2002growth}.

The CrN samples are characterized $\textit{ex situ}$ with RBS to find their stochiometry. Best fit to the RBS data of S61 reveals a 7\% N deficiency while S45 is found to be ideally stochiometric with Cr:N = 1.0:1.0. As measured by RHEED and XRD, we do not see any secondary phases in any samples; therefore, a N deficiency is attributed to N vacancies in the sample.

\subsection{Observation of {\it In-plane} Structural Transition Using RHEED}

A possible \textit{in-plane} structural transition is studied in S73 and S75 using VT-RHEED. If the room-temperature fcc crystal structure transforms into orthorhombic with a 2 degrees distortion, then the RHEED patterns may be affected by the structural changes. For example, if the entire CrN film is distorted coherently along [110] then the LT-RHEED patterns taken along [100] and [010] should be misaligned with respect to the RT-RHEED pattern, which should result in asymmetry of the streak pattern. Alternatively, if half of the sample domains of the film are stretched along [110] and the other half along [1\underline{1}0] then one would expect to see split streaks or streak broadening in RHEED patterns along [110] and [1\underline{1}0]. Such effects are not obvious in the RHEED patterns taken below T$_{N}$.

\begin{figure}[htbp]
%\centering
\includegraphics[angle=0,width= 5 in]{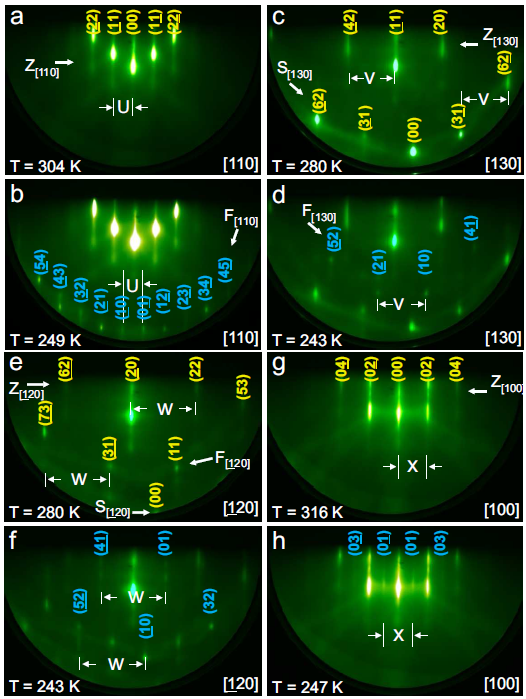}

\subfigure{\label{fig:VT-RHEED-a}}
\subfigure{\label{fig:VT-RHEED-b}}
\subfigure{\label{fig:VT-RHEED-c}}
\subfigure{\label{fig:VT-RHEED-d}}
\subfigure{\label{fig:VT-RHEED-e}}
\subfigure{\label{fig:VT-RHEED-f}}
\subfigure{\label{fig:VT-RHEED-g}}
\subfigure{\label{fig:VT-RHEED-h}}

\caption{RHEED patterns \subref{fig:VT-RHEED-a}, \subref{fig:VT-RHEED-c}, \subref{fig:VT-RHEED-e}, and \subref{fig:VT-RHEED-g} are recorded above and \subref{fig:VT-RHEED-b}, \subref{fig:VT-RHEED-d}, \subref{fig:VT-RHEED-f}, and \subref{fig:VT-RHEED-h} are recorded below the transition temperature for S73. Crystallographic direction of each pattern is given at the bottom right corner, and temperatures is shown at bottom left corner. Characteristic streak spacings along [110], [130], [$\underline{1}$20], and [100] are represented by U, V, W, and X. Zeroth, first, and second-order Laue rings are represented by Z, F, and S. Streaks/spots seen at both high and low temperature are labeled in yellow, while streaks/spots seen only at low temperature are labeled in blue.}
\label{fig:VT-RHEED}
\end{figure}

On the other hand, we do observe new streaks/spots in the RHEED patterns of S73 and S75 at 277 K and 278 K, respectively, upon cooling, as shown in Fig.~\ref{fig:VT-RHEED}. The relative brightness of the new streaks/spots increases with further decrease in the sample temperature while the overall patterns stay the same. It is an abrupt transition with a transition temperature of $\sim$ 278 K. As described in the following, the appearance of these additional streaks/spots below the transition temperature corresponds to a transition from a 1$\times$1 unit cell to a $\mathrm{\sqrt{2}\times\sqrt{2}-R45^\circ}$ unit cell (alternatively, from a primitive 1$\times$1 to a conventional 1$\times$1 unit cell).

Shown in Fig.~\ref{fig:VT-RHEED} are eight RHEED patterns of S73, where patterns shown in [\subref{fig:VT-RHEED-a}, \subref{fig:VT-RHEED-c}, \subref{fig:VT-RHEED-e}, and \subref{fig:VT-RHEED-g}] are recorded above, and patterns shown in [\subref{fig:VT-RHEED-b}, \subref{fig:VT-RHEED-d}, \subref{fig:VT-RHEED-f}, and \subref{fig:VT-RHEED-h}] are recorded below, the phase transition. The patterns are furthermore grouped in columns as follows: [110] (a,b); [130] (c,d); [$\underline{1}$20] (e,f); and [100] (g,h). These crystallographic directions can be identified by their characteristic streak spacings. For example, U is the streak spacing along [110]; and V, W, and X are larger than U by $\sqrt{5}$, $\sqrt{10}$, and $\sqrt{2}$ times, respectively. Unique streaks/spots in each RHEED pattern are marked with Miller indices which correspond to the reciprocal space map presented in Fig.~\ref{fig:LT-RHEED-Model-b}.

When we compare Fig.~\ref{fig:VT-RHEED-a} with Fig.~\ref{fig:VT-RHEED-b}, above the transition only a zeroth-order-Laue ring (Z$_{[110]}$) is visible while below the transition, in addition to Z$_{[110]}$, a first-order Laue ring (F$_{[110]}$) appears. Additionally, one can notice two points: first, the spacing between F$_{[110]}$ streaks (U) is the same as for Z$_{[110]}$; however, each F$_{[110]}$ streak occupies a 1/2-order position referenced to Z$_{[110]}$.

Similarly, in the case of [130], only zeroth-order and second-order (S$_{[130]}$) Laue rings are visible above the transition, and the streak spacing (V) in both rings is the same. Below the transition temperature, a first-order-Laue ring (F$_{[130]}$) appears midway along the vertical between Z$_{[130]}$ and S$_{[130]}$, but each spot of F$_{[130]}$ is laterally offset midway between the adjacent streaks of Z$_{[130]}$ and S$_{[130]}$, another sign of the abrupt transition.

The VT-RHEED experiments were also performed along [$\underline{1}$20] and [100] directions, and in each case 1/2-order streaks appear while cooling through the transition temperature as is clear within the zeroth-order and first-order Laue rings for [$\underline{1}$20] and within the zeroth-order Laue ring for [100].

\subsection{Model to Explain the {\it In-plane} Transition Seen in VT-RHEED}

\begin{figure}[htbp]
\centering
\includegraphics[angle=0,width= 6.4 in]{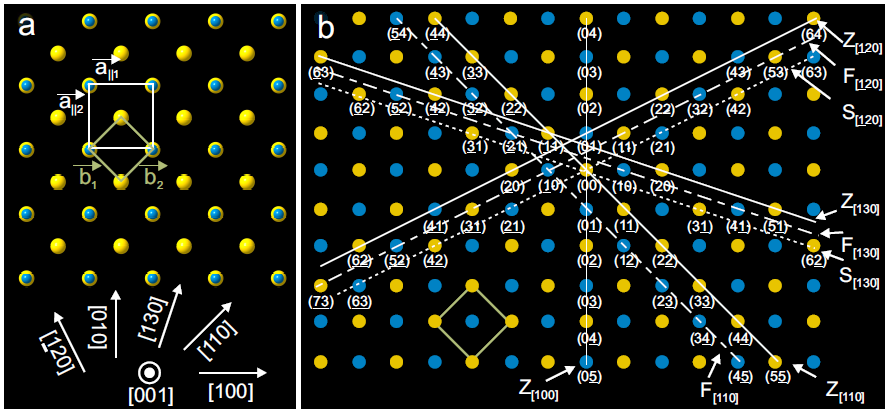}

\subfigure{\label{fig:LT-RHEED-Model-a}}
\subfigure{\label{fig:LT-RHEED-Model-b}}

\caption{\subref{fig:LT-RHEED-Model-a} Direct lattice for CrN(001), all colored (both yellow and blue) balls represent Cr atoms in the primitive 1$\times$1 structure corresponding to FCC CrN at room temperature, whereas at low temperature a superperiodicity is observed indicated by only blue balls and having a $\mathrm{\sqrt{2}\times\sqrt{2}-R45^\circ}$ unit cell relative to the 1$\times$1 primitive cell. \subref{fig:LT-RHEED-Model-b} Reciprocal lattice for CrN(001); only yellow spots are observed at RT, whereas both yellow and blue spots are observed at LT, determined by the reciprocal lattices of the 1$\times$1 and $\mathrm{\sqrt{2}\times\sqrt{2}-R45^\circ}$ unit cells. Zeroth-, first-, and second-order Laue zones are represented by Z (solid lines), F (dashed lines), and S (dotted lines), respectively. All rods corresponding to streaks/spots visible in the RHEED patterns are labeled in reciprocal space.}

\label{fig:LT-RHEED-Model}
\end{figure}

To understand the origin of the diffraction streaks/spots seen in RHEED above and below the transition temperature, a detailed model of the CrN surface is shown in Fig.~\ref{fig:LT-RHEED-Model}. Using this model, each RHEED streak/spot from the experimental patterns can be precisely accounted for. The direct lattice is shown in Fig.~\ref{fig:LT-RHEED-Model-a}, and the corresponding reciprocal lattice map is shown in Fig.~\ref{fig:LT-RHEED-Model-b}.

The basic structure for CrN(001) surface above the transition can be viewed as just a primitive 1$\times$1 square lattice (yellow atom lattice, labeled with vectors $\vec{\mathrm{b}}_{1}$ and $\vec{\mathrm{b}}_{2}$ primitive cell vectors) as shown in Fig.~\ref{fig:LT-RHEED-Model-a}. Corresponding to this 1$\times$1 primitive lattice in real space is the reciprocal space 2D square lattice (yellow spots) shown in Fig.~\ref{fig:LT-RHEED-Model-b}. This yellow spot reciprocal space map explains all the spots seen in RHEED in all 4 azimuthal directions ([110], [130], [$\underline{1}$20], and [100]) above the transition.

The b$_{1}$-b$_{2}$ primitive lattice can also be viewed as a conventional face-centered square lattice [unit vectors $\vec{\mathrm{a}}_{1}$ and $\vec{\mathrm{a}}_{2}$ shown in Fig.~\ref{fig:LT-RHEED-Model-a}] with an identical 2-atom basis. The new model below the transition can be attained simply by lifting the degeneracy between corner and face-centered atoms, leading to half yellow and half blue atoms seen in the model. And then the corresponding reciprocal lattice map consists of the previous set of (yellow) spots plus an additional new set of (blue) spots seen in Fig.~\ref{fig:LT-RHEED-Model-b}.  These new edge-center spots account for all the new spots seen in the LT-RHEED patterns. Sets of reciprocal space points along certain directions (indicated by straight lines in Fig.~\ref{fig:LT-RHEED-Model-b}) correspond to particular Laue zones seen in the RHEED patterns. Every Laue zone and every reciprocal space lattice point is labeled in the figure in a manner consistent with markings in Fig.~\ref{fig:VT-RHEED}. Based on the model shown in Fig.~\ref{fig:LT-RHEED-Model-a}, appearance of the new  F$_{[110]}$ Laue ring along [110] azimuth is associated with the lifting of the degeneracy and corresponds to a periodicity doubling (2$\times$a$_{\Vert}$/$\sqrt{2}$) along [110]. Similarly the appearance of the F$_{[130]}$ Laue zone spots for the [130] azimuth is associated with the same degeneracy lifting as for [110] and corresponds to an atomic periodicity doubling (2$\times\sqrt{2.5}$ a$_{\Vert}$) along [130]. Lastly, the appearance of the 1/2-order spots/streaks for [$\underline{1}$20] and [100] is consistent with the degeneracy lifting shown in the model of Fig.~\ref{fig:LT-RHEED-Model-a}, and corresponds to a periodicity doubling (2$\times$a$_{\Vert}$/$\sqrt{20}$) along [2$\underline{1}$0] for the  [$\underline{1}$20] azimuth and a doubling (2$\times$a$_{\Vert}$/2) along [010] for the [100] azimuth.

We find that the simple model having a cubic unit cell fully explains the RHEED spots observed at LT; however, the cause of this super-periodicity seen on the surface is unknown. It may indicate a structural distortion. Since RHEED is not spin sensitive, we cannot conclude anything about the spin ordering at the surface, and we assume that it corresponds to some kind of cooling-induced surface buckling or other structural effect. It is possible however, that the surface phase transition corresponds to the bulk phase transition for two reasons. First, the RT 1$\times$1 surface structure is bulk-like; and second, the phase transition coincides pretty closely with the one observed in neutron diffraction described below for bulk.

%%%%%%%%%%%%%%%%% VT-XRD %%%%%%%%%
\subsection{Observation of {\it Out-of-plane} Structural Transition Using VT-XRD}
\begin{figure}[htbp]
\centering
\includegraphics[angle=0,width= 6.4 in]{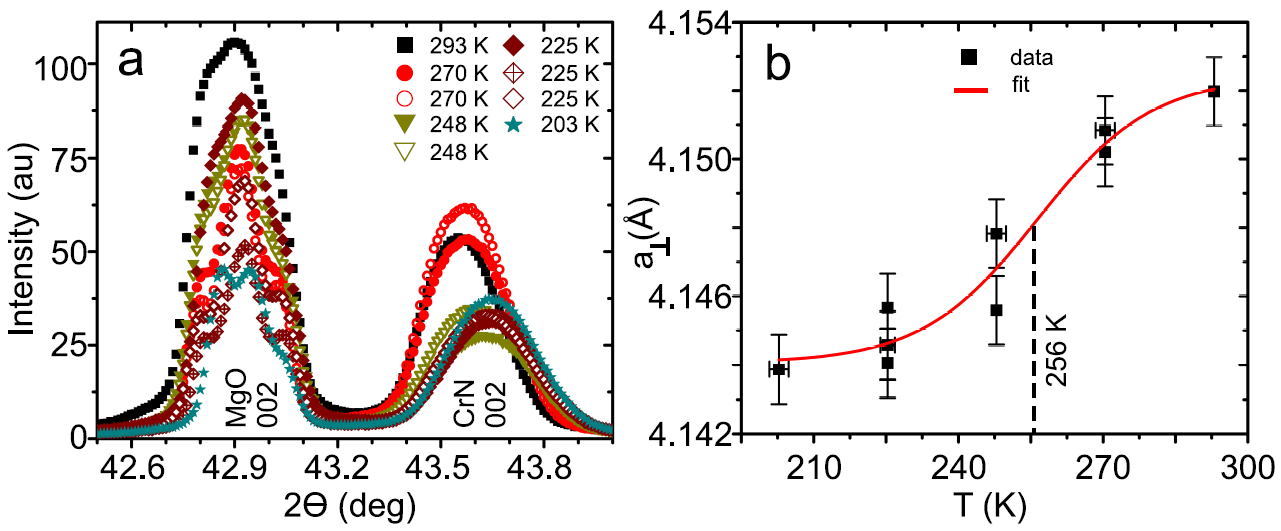}

\subfigure{\label{fig:LT-XRD-S45-a}}
\subfigure{\label{fig:LT-XRD-S45-b}}

\caption{\subref{fig:LT-XRD-S45-a} XRD spectra of S45 MgO and CrN over a range of temperature from 293 K down to 203 K; \subref{fig:LT-XRD-S45-b} {\it Out-of-plane} lattice parameter for CrN thin film as a function of temperature revealing a transition near 256 K.}

\label{fig:LT-XRD-S45}
\end{figure}

We performed VT-XRD experiments to investigate a possible \textit{out-of-plane} structural transition in the CrN films.  We cooled down sample S45 from 293 K to 203 K by flowing nitrogen vapor over the sample, and x-ray spectra were continuously recorded as the sample was either cooling down or warming up. During cooling/heating, the 002 peak of MgO as well as of CrN shift to the right/left, respectively, as shown in Fig.~\ref{fig:LT-XRD-S45-a}. Since this experiment was not done under vacuum, intensities of the peaks as well as the shape of the MgO peak change during cooling/heating, likely due to problems with icing. Despite this problem, we observed the CrN lattice constant decrease with decreasing temperature, over the range around the expected phase transition. This is seen in Fig.~\ref{fig:LT-XRD-S45-b} where  a$_{\bot}$ is plotted versus temperature. A fit to the data is obtained by using a Boltzmann equation: a$_{\bot}$(T)=4.152\AA-0.008\AA/[1+exp((T-T$_{N}$)/$\bigtriangleup$T)], which gives a transition temperature of  256 $\pm$ 6 K, which is lower than what we observed in the VT-RHEED experiment. The difference could be related to the manner in which temperature for the CrN is determined; in this experiment sample temperature is estimated from the MgO temperature determined by assuming the expected linear thermal expansion of MgO based on its linear thermal expansion coefficient ($\alpha_{MgO}$) of 9.84$\times$10$^{-6}$ K$^{-1}$ \cite{madelung2004semiconductors}; however, there was a time delay between the CrN and MgO spectra at every temperature step, which may lead to a systematic error. Therefore the result is only semi-quantitative; nonetheless, we see that the {\it out-of-plane} CrN lattice parameter appears to change in a non-linear fashion with temperature, as shown in Fig.~\ref{fig:LT-XRD-S45-b}, consistent with a phase transition. Further support for this conclusion is that if we assume a linear thermal expansion model for the CrN, then we find that the CrN thermal expansion coefficient ($\alpha_{CrN}$) is 2.9$\times$10$^{-5}$ K$^{-1}$, which is 3-4$\times$ larger than reported values 0.75-1.06$\times$10$^{-5}$ K$^{-1}$.\cite{zhang2011epitaxial, janssen2006stress,zhou2014structural, daniel2011size}

%%%%%%%%%%%% VT-ND Structural Transition %%%%%%%%%%%%%%%%
\subsection{Investigation of the Structural Transition Using VT-ND}

To more accurately investigate the structural transition in our CrN thin films, we monitored three structural peaks (111, 002, and 220, indexed relative to the pseudo-cubic unit cell) above and below the transition temperature using neutron diffraction for S61. At room temperature, all of these three peaks occur as small shoulders on the high angle side of the large 111, 002, and 220 MgO peaks since the MgO substrate thickness is substantially greater than the CrN film thickness. Three $\theta$-2$\theta$ scans through the 111 peak collected at 305, 270, and 240 K are shown in Fig. 5(a). From the 305 K and 270 K curves, we find that the position of the 111 MgO peak at 2$\theta$ = 39.6$^\circ$, in comparison to the position 39.4$^\circ$ expected for bulk MgO, does not change (note that the difference between the measured and nominal MgO 2$\theta$ values is within experimental error for the $E$ = 30.5 meV neutrons that results from the coarse resolution utilized to detect the small film reflections; uncertainties in the shift in the peak position with temperature are significantly smaller). At 305 K and 270 K, a clear shoulder is evident on the right side of the 111 MgO peak near 2$\theta$ = 40.2$^\circ$, corresponding to a $d$-spacing $d$$_{111}$ = 2.38 \AA, consistent with RT CrN lattice constants a = b = c = 4.13 \AA.

The center of the CrN shoulder at 240 K (as obtained from Gaussian fits), however, has shifted to a higher angle of 40.6$^\circ$, corresponding to a LT CrN $d$-spacing $d$$_{111}$ = 2.36 \AA. These data clearly show that there is a sudden lattice distortion (contraction) along the [111] axis of the pseudocubic cell, which is tipped at an angle of 35$^\circ$ relative to the film growth axis. These data are consistent with a bulk-like orthorhombic distortion of the CrN lattice in which the 111 peak at high temperatures splits into 201 and 011 reflections (indexed relative to the orthorhombic cell) at low temperatures. The position of the shoulder at 240 K approximately matches that expected for the bulk CrN 201 reflection (40.6$^\circ$), but the 011 orthorhombic reflection, if present, would be obscured by the 111 MgO reflection (near an angle of 39.7$^\circ$).

To determine if this structural transition is first-order, we monitored the peak intensity at 2$\theta$ = 40.2$^\circ$ as a continuous function of temperature upon heating from 240 K after cooling from room temperature. We see an abrupt change in intensity at T = 268 K, as shown in Fig. 5(b). The transition spans a range of about 12 K, but the actual transition could be sharper due to our finite temperature equilibration times (2.6 minutes per point).

Using VT-ND, we also investigated the temperature dependence of the 002 peak, as shown in Fig. 5(c). Three curves are shown, one above the transition (305 K), one near the transition (270 K), and one below the transition (240 K). A full orthorhombic distortion of the CrN (similar to what was seen for bulk powders), if it occurred, would be expected to produce only a subtle shift in the 002 peak position, from (approximately) 2$\theta$ = 46.72$^\circ$ at high temperatures to 2$\theta$ = 46.68$^\circ$ at low temperatures. This tiny shift is below the instrument resolution, and all that is seen at 240 K is a slightly decreased intensity (compared to the 270 K and 305 K data) on the right side of the MgO 002 peak near the anticipated RT CrN peak position.

To isolate any {\it in-plane} component of the lattice distortion, we performed similar neutron diffraction measurements for the 220 peak, as shown in Fig. 5(d). In this case, the anticipated RT CrN 220 peak for the curve taken at 305 K is at 68.2$^\circ$, and a shoulder near this angle is seen well separated from the MgO 220 peak (note that the fitted position of the 220 MgO peak is 2$\theta$ = 67.3$^\circ$, relative to the expected position 66.8$^\circ$ for bulk MgO). The 220 curve for 240 K (in Fig. 5d) also shows a shoulder at the same angle (Note that the 240 K data were obtained in different conditions with reduced  shielding). This indicates that the lattice constant along the [220] direction does not shift as the sample is cooled. If a full bulk-like, orthorhombic distortion occurs within the film plane, we would expect a splitting of the 220 peak at high temperatures into 020, 212 and 400 reflections (indexed relative to the orthorhombic cell) at low temperatures. The 020 (with an expected angle of 67.1$^\circ$) is obscured by the MgO substrate 220 reflection, and the 212 (with an expected angle of 68.2$^\circ$) effectively coincides with the 220 pseudocubic CrN reflection at high temperature.  At low temperature, we do not see any significant increase in scattering near the expected position of the 400 (69.4$^\circ$) though any scattering may be masked by the instrument background, which is larger due to instrument configuration differences (as noted above). So therefore, we do not find any definitive evidence of a LT lattice distortion along the {\it in-plane} direction for these CrN films, though the neutron measurements cannot rule out this possibility.

The overall conclusion from these structural studies is that a structural transition is observed for CrN in the [002] {\it out-of-plane} (by VT-XRD) and [111] directions (by VT-ND), whereas the results from VT-RHEED as well as VT-ND indicate that an expected {\it in-plane} structural distortion at low temperatures is suppressed. These data are consistent with (but not uniquely supportive of) a reduction or clamping of the {\it in-plane} lattice distortion at low temperatures due to the epitaxial constraints from the rocksalt fcc MgO substrate. The VT-ND and VT-RHEED results are consistent with a possible tetragonal type structure at low temperatures. In any case, a structural transition is definitely observed within our films.

\subsection{Measuring the Magnetic Phase Transition Using VT-ND}

Based on the work of Filippetti {\it et al.},\cite{filippetti2000magnetic} it is expected in CrN that spin ordering drives structural distortion. Therefore, we expect a connection between the observed structural transition and a magnetic phase transition in our films. Neutron diffraction is the most accurate way to probe the onset of long-range magnetic order, particularly in the case of antiferromagnetism. Sample S45 (670 nm thick) was aligned in the hhl zone, within the pseudocubic notation for the high-temperature rock-salt structure. Both the $\frac{1}{2}$$\frac{1}{2}$0 and $\frac{1}{2}$$\frac{1}{2}$1 peaks, each related to the antiferromagnetic ordering of CrN, were found below the structural transition point. Scans taken along the scattering wave vector at 100 K, namely the $\frac{h}{2}$$\frac{h}{2}$0 and $\frac{h}{2}$$\frac{h}{2}$h directions, for Figs.~\ref{fig:BT4-S45-a} and \subref{fig:BT4-S45-b} respectively, are well described by simple Gaussian line shapes when compared to the flat background at higher temperatures. The full width at half maximum (FWHM) of each peak determined from the fit is within error of the calculated resolution limit for each reflection, indicating extended correlation length and long-range magnetic order over the entirety of the sample.

The temperature dependencies of the peak intensities of the $\frac{1}{2}$$\frac{1}{2}$0 and $\frac{1}{2}$$\frac{1}{2}$1 reflections are shown in Figs.~\ref{fig:BT4-S45-c} and \subref{fig:BT4-S45-d}, respectively. The intensities of both reflections increase sharply upon cooling, consistent with a first-order phase transition. The temperature dependencies below T$_{N}$ results from a small temperature-dependent background apparent in Fig.~\ref{fig:BT4-S45-a} and \subref{fig:BT4-S45-b} rather than the sample itself. Polarized neutron beam experiments furthermore confirm that these peaks are magnetic in origin. Therefore, from the temperature dependent data, a N\'{e}el temperature is determined by fitting it with a simple order-parameter and linear background (solid line in the figure), from which we find T$_{N}$ = 280 $\pm$ 2 K. Similar measurements of the $\frac{1}{2}$ $\frac{1}{2}$0 reflection for a thicker sample (S61, 980 nm and having a 7\% nitrogen deficiency) were also performed, finding a first-order transition at a lower T$_{N}$ (270 K) than for S45.

Interestingly, the determined N\'{e}el temperature for S45 (280 $\pm$ 2 K) is very close to the structural transition temperature determined from VT-RHEED measurements (278 K), although those samples were much thinner (S73/150 nm and S75/37 nm). This suggests a possible close correspondence between magnetic and structural transitions in these CrN films. However, the correspondence is not as close when comparing the N\'{e}el temperature (280 K) with the structural transition temperature determined for the same sample by means of VT-XRD (256 K), possibly due to temperature equilibration issues in the VT-XRD measurement. On the other hand, the reduced N\'{e}el temperature (270 K) for 7\% nitrogen deficient S61 matches the transition temperature obtained for the same sample using VT-ND (268 K). In any case, it is clear that these CrN films exhibit both magnetic and structural transitions, and over similar temperature ranges.

%%%%%%%%%%% Theory and AFM model %%%%%%%%%%%%%%
\subsection{Theoretical Calculations of Cubic, Tetragonal, and Orthorhombic Models}

When we compare our experimental results to the Corliss model (AFM-[110]$_2$), which was also investigated by Filippetti \textit{et al.},\cite{corliss1960antiferromagnetic,filippetti2000magnetic} as shown in Fig.~\ref{fig:aFM-Model-a} and \subref{fig:aFM-Model-b}, we find excellent agreement magnetically but structurally the picture appears more complicated. Certainly, the Corliss model with its alternating double-layer ferromagnetic sheets (Type 4 antiferromagnetism according to the Cox fcc classification system) gives rise to $\frac{1}{2}$$\frac{1}{2}$0 and $\frac{1}{2}$$\frac{1}{2}$1 magnetic peaks which we observed, and yet the RHEED did not observe any {\it in-plane} distortion expected for the orthorhombic model, consistent with the VT-ND results which did not observe any changes in the 220 peak position after cooling the sample. On the other hand, the 111 peak was observed to shift, exhibiting a first-order phase transition versus temperature. These results suggested the possibility of other models which could explain the magnetism while also giving a better agreement structurally. For example, the LT-RHEED data suggest a square lattice unit cell, and given the results from VT-XRD and VT-ND, it would therefore make sense to consider both cubic and tetragonal AFM models, in addition to the AFM-[110]$_2$ model. Two such models are presented in Fig.~\ref{fig:aFM-Model-c} (AFM-cubic, Type 1 AFM) and Fig.~\ref{fig:aFM-Model-e} (AFM-tetragonal, Type 3 AFM). Additional motivation to consider other models comes from the theoretical work of Liangcai \textit{et al.} who showed that the AFM-[110]$_2$ model in bulk CrN has lower total energy compared to the AFM-cubic model by only 0.06 eV/atom or 0.04 eV/atom depending on the particular method,\cite{zhou2014structural} which is not a big difference, and therefore epitaxial constraints as well as strain in actual films could be enough to change the sign of the inequality.

We therefore performed numerical calculations for non-magnetic, ferromagnetic, and antiferromagnetic ordering in the cases of cubic, tetragonal, and orthorhombic models using GGA (generalized gradient approximation) and LDA+U (local density approximation + Hubbard correction, with 3 eV $\leq$U$\leq$ 5 eV). Our models are divided into two categories based on the number of monolayers (ML's). The models with $<$ 9 ML's (surface) give a lattice constant of 4.140 \AA, whereas models with $\geq$ 9 ML's (bulk) are relatively relaxed and give a lattice constant of 4.145 \AA.

For the AFM-cubic model, both the surface and bulk models show metallic behavior. Whereas the surface model of the AFM-tetragonal shows metallic behavior, while the bulk model shows semiconducting behavior with a band gap of 0.03 eV. Similarly, for the AFM-[110]$_2$ model, we find metallic behavior for the surface model but semiconducting behavior with a band gap of 0.16 eV for the bulk model. These findings are comparable to the results of Herwadkar and Lambrecht who obtained a gap of 0.4 eV for U = 3 eV in LSDA+U calculations \cite{herwadkar2009electronic} (see also Supplementary material, figure 3). The fact that our LT-STM data is consistent with metallic behavior, which is also consistent with the results of Constantin {\it et al.'s} LT resistivity measurements for similarly grown samples \cite{constantin2004metal}, is therefore not obviously in the best agreement with the existence of a finite gap as seen in the AFM-[110]$_2$ model, as compared to a zero gap model such as AFM-cubic. However, as discussed in detail by Herwadkar and Lambrecht, this LT metallic behavior could arise from exceeding a critical electron density owing to N vacancies acting as donors.

We should emphasize here that for each model we began our calculations from a perfect cubic structure and the models were allowed to relax in all directions. The AFM-[110]$_2$ model shows a shear distortion similar to what was observed by Corliss {\it et al.} and Filippetti \textit{et al.},\cite{corliss1960antiferromagnetic,filippetti2000magnetic}, while the AFM-cubic model shows no distortion [shown in Figs.~\ref{fig:aFM-Model-c} and \subref{fig:aFM-Model-d}], and the AFM-tetragonal model [shown in Figs.~\ref{fig:aFM-Model-e} and \subref{fig:aFM-Model-f}] shows a contraction of -0.01855 \AA\ in a$_{\bot}$. The shear distortion only occurs if we choose the Corliss (AFM-[110]$_2$) magnetic model (alternating double ferromagnetic layers).

Concerning the cubic (Type 1 AFM) and tetragonal (Type 3 AFM) models, these both have very similar spin ordering. However, in the AFM-cubic model, single ferromagnetic planes alternate along [010]; while in the AFM-tetragonal model, spin rows appear rotated by 90$^\circ$ after every 1 ML when viewed along [001] as shown in Fig.~\ref{fig:aFM-Model-e}, and spin rows in the layers stacked vertically along [001] shift by a/2 along [100] after every 2 ML as shown in Fig.~\ref{fig:aFM-Model-f}.

To determine the best model, we employed the minimum energy criteria. Energies of the ferromagnetic and non-magnetic models are both higher than all the AFM models; therefore, they are not shown in this paper. The energies of the AFM-cubic and AFM-tetragonal compared to the AFM-[110]$_2$ model are found to be 0.040 eV/atom higher (consistent with the 0.04-0.05 eV/atom reported by Liangcai \textit{et al.} \cite{zhou2014structural}) and 0.023 eV/atom higher, respectively. Therefore, we confirmed that not only the AFM-cubic but also the AFM-tetragonal models are energetically less favorable compared to the AFM-[110]$_2$ model.

Most importantly, it must be realized that the observed $\frac{1}{2}$$\frac{1}{2}$0 and $\frac{1}{2}$$\frac{1}{2}$1 magnetic peaks seen in LT-ND at low temperature are forbidden in the cases of both Type 1 and Type 3 FCC AFM ordering. This can easily be realized if one simply considers that for a 1/2-order magnetic peak to exist in reciprocal space, it requires a spatial doubling of the magnetic unit cell. Such a doubling occurs in the case of the Type 4 AFM-[110]$_2$ model as can be seen in Fig.~\ref{fig:aFM-Model-a}. This doubling also occurs along one axis only in the case of the AFM-tetragonal model; however, this will not give rise to $\frac{1}{2}$$\frac{1}{2}$0 and $\frac{1}{2}$$\frac{1}{2}$1 peaks since we require spatial doubling along not one but {\it two} orthogonal axes. Therefore, neither the AFM-cubic nor the AFM-tetragonal models can explain the observed magnetic ordering in these films. It is then surprising that although the LT-RHEED data suggests a possible cubic/tetragonal structural model, and the LT-ND data do not uniquely support an orthorhombic structural model, nonetheless only the Type 4 AFM-[110]$_2$ model can explain the magnetic results.

It remains to investigate the (001) surface of CrN using atomic resolution STM and spin-resolved STM in order to probe the surface structure and its spin magnetic ordering and compare that with the results from measurements reported here. Such future studies could address possible differences between surface and bulk properties, and between structural and magnetic behaviors. Really thin films could differ from thicker ones and even possibly stabilize the AFM-tetragonal state, and in any case STM measurements could prove essential in order to fully understand the CrN system.

%%%%%%%%%%%%%%%%%%%%%%%%%%%%%%%%%%%%%%%%%%%%%%%%%%%%%%%%%%%%%%%%
\section{Conclusions}

Variable temperature RHEED reveals a clear structural transition with a transition temperature of 277-278 K in high-quality, MBE grown CrN/MgO(001) films as thin as 37 nm. The epitaxial samples were also investigated at NCNR using a triple axis spectrometer, and the results confirm a magnetic phase transition at 280 K and at 270 K (for a nitrogen deficient film). These observed structural and magnetic transitions may be correlated with an electronic phase transition observed by Constantin \textit{et al.} and occurring near 280 K (T increasing) or 260 K (T decreasing) for similarly grown films.\cite{constantin2004metal}

First-principles theoretical calculations were employed to investigate possible structural/magnetic models including non-magnetic, ferromagnetic, and antiferromagnetic models. The three antiferromagnetic models: AFM-[110]$_2$ (Corliss model), AFM-cubic, and AFM-tetragonal, are energetically better than non-magnetic or ferromagnetic, with the AFM-[110]$_2$ model being the energetically most favorable model. Whereas the structural symmetry observed by LT-RHEED suggests a cubic or tetragonal model, and although VT-XRD and VT-ND do not uniquely determine a full orthorhombic model, required magnetic symmetry to explain the observed $\frac{1}{2}$$\frac{1}{2}$0 and $\frac{1}{2}$$\frac{1}{2}$1 magnetic peaks seen in ND precludes the possibility of having either the AFM-cubic or AFM-tetragonal model. It is therefore somewhat surprising that despite the possibility that the film is prevented from undergoing a full orthorhombic distortion, possibly due to the epitaxial constraint of the cubic MgO substrate, nonetheless the double-layer AFM-[110]$_2$ magnetism is still established.

%%%%%%%%%%%%%%%%%%%%%%%%%%%%%%%%%%%%%%%%%%%%%%%%%%%%%%%%%%%%%%
\newpage
\section{acknowledgments}

Research supported by the U.S. Department of Energy, Office of Basic Energy Sciences, Division of Materials Sciences and Engineering under Award \# DE-FG02-06ER46317. The authors thank Joseph P. Corbett for useful discussion and help in figure preparation. The authors would also like to acknowledge Martin E. Kordesch for back-coating MgO(001) substrates with titanium. The authors also acknowledge WSxM software for STM image processing \cite{horcas2007wsxm}. N.T. thanks Conacyt project 164485 and DGAPA-UNAM project IN100516 for partial financial support. Calculations were performed in the DGCTIC-UNAM supercomputing center, project SC16-1-IG-31.

%%%%%%%%%%%%%%%%%%%%%%%%%%%%%%%%%%%%%%%%%%%%%%%%%%%%%%%%%%%%%%%%%%%%%

%\bibliography{references1}

%%%%%%%%%%%%%%%%%%%%%%%%%%%%%%%%%%%%%%%%%%%%%%%%%%%%%%%%%%%%%%%%%%%%

\newpage
\clearpage

%%%%%%%%%%%%%%%%%%%%%%%%%%%%%%%%%%%%%%%%%%%%%%%%%
\newpage
\clearpage

%%%%%%%%%%%%%%%%%%%%%%%%%%%%%%%%%%%%%%%%%%%%%%%
\newpage
\clearpage
%%%%%%%%%%%%%%%%%%%%%%%%%%%%%%%%%%%%%%%%%%%%%%%%%
\newpage

\clearpage
%%%%%%%%%%%%%%%%%%%%%%%%%%%%%%%%%%%%%%%%%%%%%%%%%
\newpage
\begin{figure}[htbp]
\centering
\includegraphics[angle=0,width= 6.4 in]{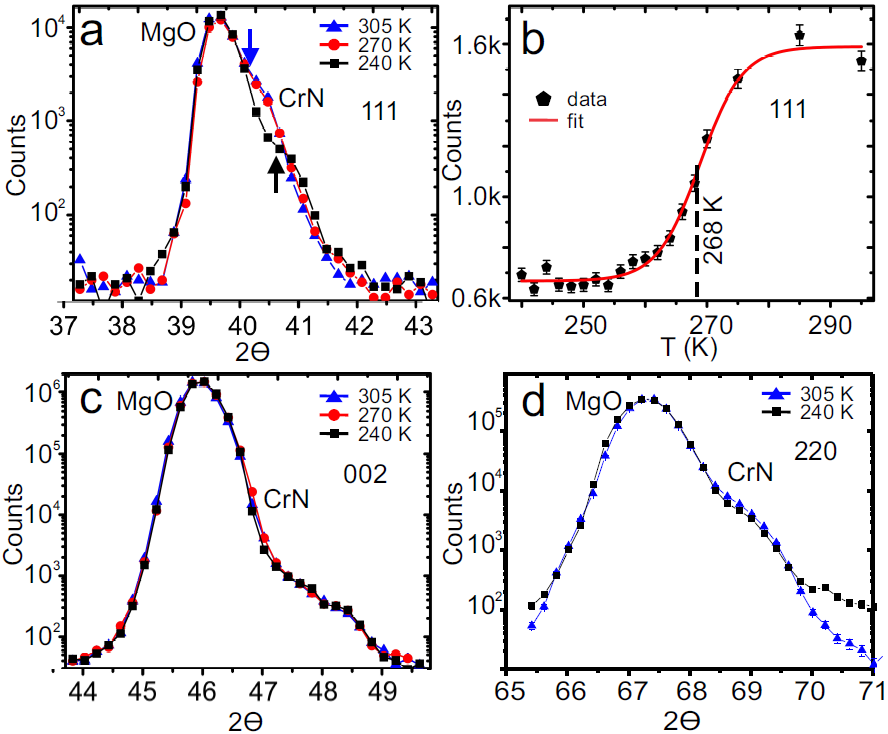}

\subfigure{\label{fig:LT-ND-S61-a}}
\subfigure{\label{fig:LT-ND-S61-b}}
\subfigure{\label{fig:LT-ND-S61-c}}
\subfigure{\label{fig:LT-ND-S61-d}}

\caption{\subref{fig:LT-ND-S61-a} The CrN 111 peak (for S61), a shoulder of the MgO 111 peak, is clearly shifted from 2$\theta$ = 40.2$^\circ$ at 305 K and 270 K, to 40.6$^\circ$ at 240 K. \subref{fig:LT-ND-S61-b} peak intensity at 40.2$^\circ$ versus temperature while heating showing a first-order transition at 268 K; \subref{fig:LT-ND-S61-c} the CrN 002 peak (for S61), which should be located on the right shoulder of the MgO 002 peak at 46.7$^\circ$, shows no significant difference from 305 K to 270 K, but at 240 K a slightly decreased intensity is observed; \subref{fig:LT-ND-S61-d} the CrN 220 peak (for S61), expected at 68.2$^\circ$, appears as a clear shoulder of the MgO 220 peak, and no change in position is observed between 305 K and 240 K. Note that the 240 K data was obtained in different conditions with reduced shielding which gives rise to higher count rates and greater background levels. For comparison purposes, the intensity has been scaled such that the MgO peak intensity matches that obtained in the 305 K scan.}

\label{fig:LT-ND-S61}
\end{figure}

\clearpage
%%%%%%%%%%%%%%%%%%%%%%%%%%%%%%%%%%%%%%%%%%%%%%%%%
\newpage
\begin{figure}[htbp]
\centering
\includegraphics[angle=0,width= 6.0 in]{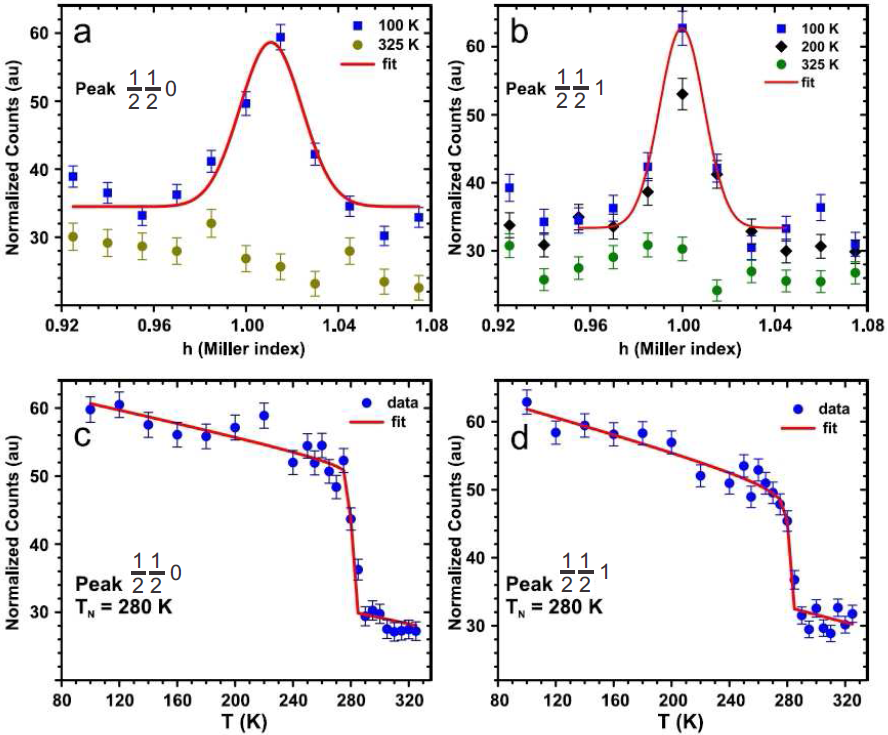}

\subfigure{\label{fig:BT4-S45-a}}
\subfigure{\label{fig:BT4-S45-b}}
\subfigure{\label{fig:BT4-S45-c}}
\subfigure{\label{fig:BT4-S45-d}}

\caption{\subref{fig:BT4-S45-a} Two $\frac{h}{2}$$\frac{h}{2}$0 scans of the $\frac{1}{2}$$\frac{1}{2}$0 region, one at 100 K showing a magnetic peak close to h = 1.01, and another at 325 K showing no peak. \subref{fig:BT4-S45-b} three $\frac{h}{2}$$\frac{h}{2}$h scans of the $\frac{1}{2}$$\frac{1}{2}$1 region, showing a magnetic peak at h = 1.00 for 100 K and 200 K, but no peak for 325 K. \subref{fig:BT4-S45-c}/\subref{fig:BT4-S45-d} Plots of the intensities taken at the peak centers versus temperature for the $\frac{1}{2}$$\frac{1}{2}$0 and $\frac{1}{2}$$\frac{1}{2}$1 peaks, covering the range from 100 K to 325 K. Clear first-order phase transitions are observed with N\'{e}el temperatures of 280 K $\pm$ 1 K. Sample used: CrN S45.}
\label{fig:Fig-4-S45}
\end{figure}
\clearpage

%%%%%%%%%%%%%%%%%%%%%%%%%%%%%%%%%%%%%%%%%%%%%%%
\newpage
\begin{figure}[htbp]
\centering
\includegraphics[angle=0,width= 6.4 in]{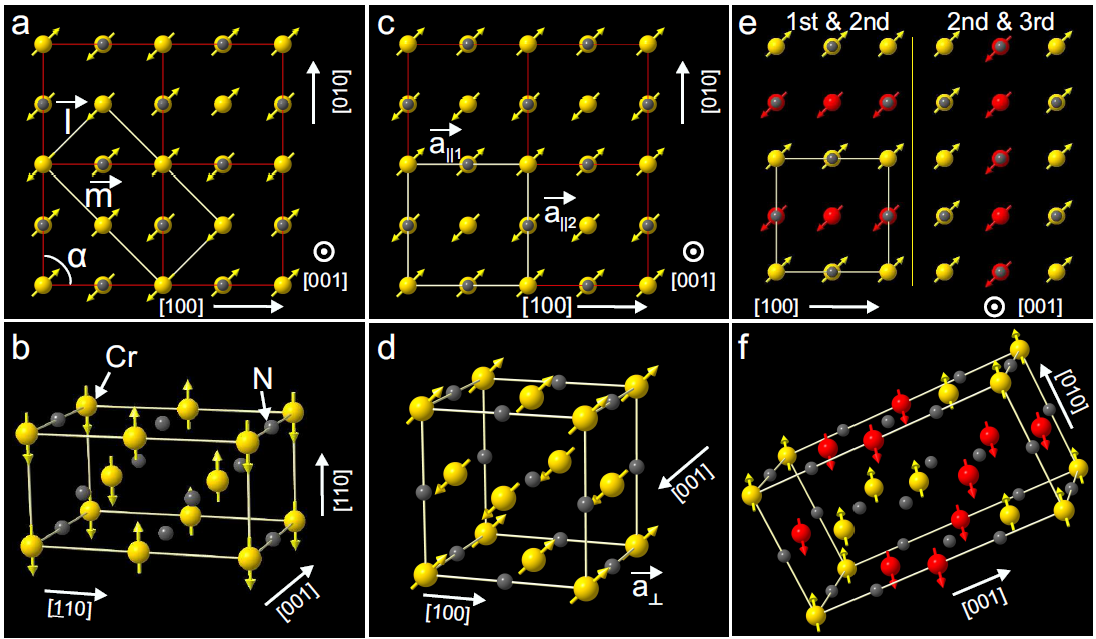}

\subfigure{\label{fig:aFM-Model-a}}
\subfigure{\label{fig:aFM-Model-b}}
\subfigure{\label{fig:aFM-Model-c}}
\subfigure{\label{fig:aFM-Model-d}}
\subfigure{\label{fig:aFM-Model-e}}
\subfigure{\label{fig:aFM-Model-f}}

\caption{\subref{fig:aFM-Model-a} Corliss structural/magnetic model (AFM-[110]$_2$) for CrN at LT involving a shear distortion with $\alpha$ $\approx$ 88$^\circ$. Double ferromagnetic (110) sheets alternate spin direction along [$\bar{1}$10]; four pseudocubic unit cells are shown along with a 45$^\circ$-rotated, orthorhombic unit cell; \subref{fig:aFM-Model-b} 3D rendering of the Corliss model orthorhombic unit cell; \subref{fig:aFM-Model-c} AFM-cubic model in which single ferromagnetic layers alternate along [010] while spins at the (001) surface form a $\mathrm{\sqrt{2}\times\sqrt{2}-R45^\circ}$ unit cell; \subref{fig:aFM-Model-d} 3D view of the AFM-cubic unit cell; \subref{fig:aFM-Model-e} top view model including three layers of the AFM-tetragonal model; spin rows in the 1st and 2nd layers are parallel to [100], while spin rows in the 2nd and 3rd layers are parallel to [010]; \subref{fig:aFM-Model-f} 3D-rendered view of the AFM-tetragonal model.}
\label{fig:aFM-Model}
\end{figure}
\clearpage

\end{document}